\documentclass[preprint,nofootinbib,amsmath,prd,aps,superscriptaddress,tightenlines,12pt]{revtex4}
\usepackage{graphicx}
\usepackage{graphics}
\usepackage{amsmath}
\usepackage{hyperref}
\usepackage{xcolor}
\usepackage{xspace}
\usepackage{subfigure}

\def\lsim{\mathrel{\rlap{\lower4pt\hbox{\hskip1pt$\sim$}}
    \raise1pt\hbox{$<$}}}                
\def\gsim{\mathrel{\rlap{\lower4pt\hbox{\hskip1pt$\sim$}}
    \raise1pt\hbox{$>$}}}                

\def\OMIT#1{}

\newcommand{\be}{\begin{eqnarray}}
\newcommand{\ee}{\end{eqnarray}}

\newcommand{\nn}{\nonumber}

\newcommand{\bea}{\begin{eqnarray}}
\newcommand{\eea}{\end{eqnarray}}

\newcommand{\pTcut}{p_T^{\rm cut}}

\def\lsim{\mathrel{\rlap{\lower4pt\hbox{\hskip1pt$\sim$}}
    \raise1pt\hbox{$<$}}}                
\def\gsim{\mathrel{\rlap{\lower4pt\hbox{\hskip1pt$\sim$}}
    \raise1pt\hbox{$>$}}}                

\def\OMIT#1{}

\begin{document}

\setlength\baselineskip{17pt}

\begin{flushright}
\vbox{
\begin{tabular}{l}
\end{tabular}
}
\end{flushright}
\vspace{0.1cm}


\title{\bf Jet vetoes for Higgs production at future hadron colliders}

\vspace*{1cm}

\author{Radja Boughezal}
\email[]{rboughezal@anl.gov}
\affiliation{High Energy Physics Division, Argonne National Laboratory, Argonne, IL 60439, USA} 
\author{Christfried Focke}
\email[]{christfried.focke@gmail.com}
\affiliation{Department of Physics \& Astronomy, Northwestern University, Evanston, IL 60208, USA}
\author{Ye Li}
\email[]{yli@slac.stanford.edu}
\affiliation{SLAC National Accelerator Laboratory, Stanford University, Stanford, CA 94309, USA} 
\author{Xiaohui Liu}
\email[]{xiaohui.liu@northwestern.edu}
\affiliation{High Energy Physics Division, Argonne National Laboratory, Argonne, IL 60439, USA} 
\affiliation{Department of Physics \& Astronomy, Northwestern University, Evanston, IL 60208, USA}


  \vspace*{0.3cm}

\begin{abstract}
  \vspace{0.5cm}

We study Higgs boson production in exclusive jet bins at possible future 33 and 100 TeV proton-proton colliders.  We compare the cross sections obtained using fixed-order perturbation theory with those obtained by also resuming large logarithms induced by the jet-binning in the gluon-fusion and associated production channels.  The central values obtained by the best-available fixed-order predictions differ by $10-20\%$ from those obtained after including resummation over the majority of phase-space regions considered. Additionally, including the resummation dramatically reduces the residual scale variation in these regions, often by a factor of two or more.  We further show that in several new kinematic regimes that can be explored at these high-energy machines, the inclusion of resummation improvement is mandatory.

\end{abstract}

\maketitle

\section{Introduction}
\label{sec:intro}

The ATLAS and CMS experiments have discovered a new boson, and the measured properties of this state are so far consistent with those of the Standard Model (SM) Higgs boson~\cite{Aad:2012tfa, Chatrchyan:2012ufa}. One of the major goals of the LHC in the next few years is the precision study of the properties of the Higgs particle. This requires precise experimental measurements coupled with an accurate modeling of the signal and background on the theory side.  Unfortunately the discovery of the Higgs does not answer all the open questions; for example, whether there is a deeper principle underlying the Higgs mechanism we have found, and what are the origins of dark matter and the matter-antimatter asymmetry in the universe.  The discovery of the Higgs gives us a place to search for answers to these questions.  However, their answers may require energies beyond those accessible at the LHC.  In particular, precise tests of Higgs couplings to Standard Model particles, the self-coupling of the Higgs and the structure of its potential may require a future higher-energy proton-proton collider.  The past year has seen a growing interest in the physics of a possible Future Hadron Collider (FHC).  It is a candidate to continue exploration of the energy frontier once the LHC program is complete  in roughly 20 years from now~\cite{FHC}.

It is clear from the experience so far at the LHC that precision SM theory will continue to be crucial in supporting and guiding any program at a FHC.  Predictions will need to be made for both inclusive cross sections as well as cross sections with experimental selection cuts.  The higher energies at proposed future machines allow for increasingly high-energy scattering events, which pose interesting new challenges for precision QCD theory.  It is almost certain that many of the experimental cuts used at the LHC to reduce backgrounds will be required at a FHC.  Since higher scattering energies will be probed, this will lead to increasingly stringent cuts on QCD radiation that produces spurious jets in addition to those contained in signal processes.  The large logarithms in question take the form $L\,=\, \ln\left(  Q/p_{cut}\right)$, where Q is the hard sale of the considered process and $p_{cut}$ is the scale of the cut on QCD radiation.  At a FHC, $Q$ will significantly increase.  Although $p_{cut}$ will likely also increase due to increased soft hadronic activity accompanying each event, there will be a desire to keep this low in order to reduce backgrounds.  The role of resummation of these large logarithms $L$ will become more central at future machines.  

Our goal in this manuscript is to consider the effect of such logarithms in future hadronic collisions by studying example processes.  We consider two examples in Higgs physics, as it is an area that will remain vitally important in future experiments, and also because it famously requires significant cuts to separate signal from background.  One example is the $H\to WW^* \to \ell^+ \ell^- \nu \bar{\nu}$ analysis, where the events are binned by exclusive jet multiplicity~\cite{Aad:2012tfa,Aad:2012uub,Chatrchyan:2012ufa,Chatrchyan:2012ty}.  The power of the analysis comes from separating the 0-jet and 1-jet bins from the inclusive 2-jet bin, where the $t\bar{t}$ background contamination is large.  Binning by jet multiplicity allows the gluon-fusion production of the Higgs to be probed in the 0-jet and 1-jet bins, and the vector-boson fusion mode to be studied by using different cuts in the 2-jet bin.  It is well known that predictions in fixed-order perturbation theory for this process 
 can suffer from large uncertainties when selection cuts are applied due in part to unresummed logarithms involving the relevant scales in the process~\cite{Berger:2010xi,Stewart:2011cf,Gangal:2013nxa,Stewart:2013faa,Banfi:2012yh, Banfi:2012jm, Becher:2012qa, Banfi:2013eda, Becher:2013xia, Tackmann:2012bt, Liu:2012sz,Liu:2013hba,Boughezal:2013oha}.  For the 0-jet bin the relevant scales are $Q=m_H$ and $p_{cut} = \pTcut$, while for the 1-jet bin we instead have  $Q=p_{TJ}$.  By resumming these logarithms to all orders, the perturbative uncertainties can be considerably reduced.  We can see clearly that at a FHC, much larger $p_{TJ}$ can be probed, leading to larger logarithmic corrections.  

A second example is the case of the $VH$ process in the boosted regime, where $V=W,Z$.  This process has been suggested as a possible place to measure the coupling of the bottom quark to the Higgs~\cite{Butterworth:2008iy}.  As the transverse momenta of the $V$ and $H$ are increased, the $H \to \bar{b}b$ decay becomes collimated, creating a ``fat jet" distinct from those produced by QCD.  In order to reduce the $t\bar{t}$ background, a jet veto is again imposed.  For high transverse momenta, logarithms of the transverse momenta over the veto scale become large.  There is already evidence for the onset of these logarithms in the fixed-order perturbation theory for the kinematic region at the LHC.  The cross section increases by $+30\%$ when going from leading-order (LO) to next-to-next-to-leading order (NNLO), while after the imposition of a jet veto the cross section decreases by more than a factor of two at high transverse momentum when going from LO to NNLO~\cite{Ferrera:2011bk}.  This change is even larger with the increased phase space at a FHC~\cite{Campbell:2013qaa}.

We study here these two example processes in 33, and 100 TeV proton-proton collisions.  We compare the results of fixed-order perturbation with those from resummation-improved perturbation theory.  Our results use the best available theoretical predictions; at fixed order we use the NNLO Higgs+0-jet cross section, NLO Higgs+1-jet result, and the NNLO $VH$ cross section.  We use the highest resummation accuracy available for each process.  Our theoretical framework for performing the resummation is soft-collinear effective theory (SCET)~\cite{Bauer:2000ew,Bauer:2000yr,Bauer:2001ct,Bauer:2001yt,Bauer:2002nz}.  We carefully consider the theoretical uncertainties affecting each process, using the standard Stewart-Tackmann prescription at fixed-order~\cite{Stewart:2011cf}, and the established method for estimating theoretical uncertainties in SCET~\cite{Abbate:2010xh,Ligeti:2008ac}.

Our paper is organized as follows. In Section~\ref{sec:formalism} we briefly describe the theoretical formalism used in this paper.  In Section~\ref{sec:numerics}
we show numerical results for Higgs production in gluon fusion in association with zero, one and two or more jets, as well as for $W^{+}H$ production in the 0-jet bin. Finally, we conclude in Section~\ref{sec:conclusion}.
 
\section{Theoretical formalism}
\label{sec:formalism}

We present here a brief review of the theoretical formalism used to obtain the results presented in our paper.  We attempt to give simple and intuitive explanations of the resummations we have performed, since the technical details have been extensively documented elsewhere.  We present only the schematic formulae here, and refer the reader to the relevant papers for more detail.  We accomplish the resummation of these large logarithms using soft-collinear effective theory (SCET)~\cite{Bauer:2000ew,Bauer:2000yr,Bauer:2001ct,Bauer:2001yt,Bauer:2002nz}.  The application of this effective theory to the problem of gluon-fusion Higgs production in the 0-jet bin has been discussed in detail in the literature~\cite{Becher:2012qa,Tackmann:2012bt,Becher:2013xia,Stewart:2013faa}; we review the salient details here to explain our approach.  For work on performing the resummation of jet-veto logarithms in the 0-jet bin using the traditional QCD approach, we refer the reader to Refs.~\cite{Banfi:2012yh,Banfi:2012jm}.  The resummation of the large logarithms in the 1-jet bin has been studied using SCET~\cite{Liu:2012sz,Liu:2013hba,Boughezal:2013oha}, as has the consistent combination of the 0-jet and 1-jet bins~\cite{Boughezal:2013oha}.  The resummation of the 0-jet bin for $VH$ production was considered in Refs.~\cite{Shao:2013uba,Li:2014ria}.  For the $H+1$-jet and $VH$ processes, we refer the reader to these papers for all technical details and formulae.
	
\subsection{Pedagogical introduction to resummation using SCET}

We begin by discussing the structure of the perturbative series for the Higgs+0-jet cross section as an example.  The fixed-order cross section 
takes the following form:
\be
\sigma_0^{FO}(\pTcut) \label{FO} = \sigma_0^{(0)}\,\sum_n \sum_{m<2n}\alpha_s^n c_{n}^{m}L^{2n-m} + \sigma_{\rm ns} 
= \sigma_{sing.} + \sigma_{\rm ns}\,,
\ee
where $L = \log({\pTcut/m_H})$.  We have introduced the notation $\sigma_{\rm ns}$ for the non-singular component of the cross section, and 
$\sigma_{sing.}$ for the singular part containing the large logarithms.  As $\pTcut \to 0$, $\sigma_{\rm ns} \to 0$.  At NLO in $\alpha_s$, this cross section becomes
\bea\label{FOa}
\sigma^{\rm FO}_0(\pTcut) &= &\sigma_0^{(0)}\left[1 + \alpha_s (c^{(1)}_2L^2 + c^{(1)}_1L + c_0^{(1)}) \right] + \sigma_{\rm ns}
+{\cal O}(\alpha_s^2)  \,.
\eea
When $\pTcut \ll m_H$, $\alpha_s L^2 \sim 1$. Therefore, the perturbative series becomes
unstable, potentially resulting in both unreliable predictions and large scale uncertainties.  To retain predictive power 
we must resum the large logarithms to all orders in the strong coupling constant:
\bea\label{sumFO}
\sigma^{{\rm resum}+{\rm FO}}_0(\pTcut) &=& \sigma_0^{(0)}\left[1 + \alpha_s  c_0^{(1)} \right] 
\exp\left[\alpha_s \left( c^{(1)}_2L^2 + c^{(1)}_1L  \right) \right]
+ \left\{ \sigma_{\rm ns}
+{\cal O}(\alpha_s^2) \right\}  + {\cal O}( \alpha_s^2 L) \nn \\
&=& \sigma_{res.} + \sigma_{\rm ns}
\,.
\eea

Very roughly, the factorization theorem established using SCET separates the large jet-veto logarithm in the following way:
\bea\label{F1}
L^2 = \log^2 \frac{m_H}{\mu} + 2 \log \frac{\pTcut}{\mu} \log \frac{\nu}{m_H} \,
+ \log \frac{\pTcut}{\mu} \log \frac{\mu \pTcut}{\nu^2}\,.
\eea
On the right hand side of the equation, the logarithm has been ``factorized" into three
terms, none of which depends on the large kinematic ratio $\pTcut/m_H$. 
The only possible large logarthmic structures appear as the ratio of the kinematics 
and the fictitious scales introduced: $m_H/\mu$, $\pTcut\mu$, $\nu/m_H$ and $\mu/\nu$.  The singular cross section 
for the 0-jet cross section is then written as
\bea\label{fact.}
\sigma_{sing.} = \sigma_0^{(0)} H(m_H,\mu)  B(\pTcut,m_H,\mu,\nu) B(\pTcut,m_H,\mu,\nu) 
S(\pTcut,\mu,\nu) \,,
\eea
where just like the naive separation in Eq.~(\ref{F1}), in each function no
logarithm depending on $\pTcut/m_H$ occurs.  Here, $H$ is the hard function including the virtual corrections,
$B$ is the beam function including collinear radiation in the $\pm \hat{z}$ directions, and $S$ is the soft function describing soft radiation.  Decomposing the momenta according to their light-cone components as $p \sim (p^+, p^-, p_{\perp})$, the 
beam-collinear radiation has the scaling $p_{c} \sim (m_H,(\pTcut)^2/m_H, \pTcut)$, while the soft radiation has the scaling $p_s \sim (\pTcut,\pTcut,\pTcut)$.  The other beam-collinear radiation $p_{\bar{c}}$ has a similar scaling to $p_c$ with the plus and minus light-cone components switched.  The following logarithms appear in each of these functions:
\bea\label{logs}
&&H(m_H,\mu)  \supset \log \frac{m_H}{\mu}  \,, \nn \\
&&B(\pTcut,m_H,\mu,\nu) \supset \log \frac{\pTcut}{\mu}, \log \frac{\nu}{m_H}  \,, \nn \\
&&S(\pTcut,\mu,\nu) \supset  \log \frac{\pTcut}{\mu}, \log \frac{\mu \pTcut}{\nu^2} \, . 
\eea

In order to derive this factorization formula, the full QCD cross section is expanded around the various soft and collinear limits.  In doing so divergences are introduced in each of the separate functions $H$, $B$, and $S$.  These are interpreted as ultraviolat divergences in the effective theory.  They are regulated using dimensional regularization, leading to the appearance of the usual dimensional regularization mass parameter $\mu$.  In this case there are additional rapidity divergences~\cite{Chiu:2012ir} that necessitate the appearance of an additional mass parameter $\nu$.  These divergences cancel when the full cross sections is formed.  However, they lead to renormalization-group equations satisfied by each of the separate functions in the effective theory:
\bea
\mu\frac{\mathrm{d} F }{\mathrm{d} \mu}= \gamma_{\mu,F} F \,, \quad \quad
\nu\frac{\mathrm{d} F }{\mathrm{d} \nu}= \gamma_{\nu,F} F \,,
\eea
with $F = H,B,S$.  The RG-equations allow us to resum the logarithms to all orders
in Eq.~(\ref{logs}), which gives
\bea \label{Ufac}
F(\mu,\nu) = U_F(\mu,\nu,\mu_F,\nu_F) F(\mu_F,\nu_F) \,,
\eea
where $U_F$ is the evolution kernel for function $F$, taking $ F$ from its natural scales
$\mu_F$ or $\nu_F$ to the scale $\mu$ or $\nu$ to evaluate the cross section. The scales $\mu_F$ and $\nu_F$ are chosen in such way that the perturbative calculation
of the function $F$ is justified. Therefore, the choice $\mu_F$ and $\nu_F$ will tend to
minimize the logarithms that occur inside each function. From Eq.~(\ref{logs}), we deduce that
\bea
\mu_H \sim m_H\,, \quad
\mu_B \sim \pTcut\,, \nu_B \sim m_H \,, \quad
\mu_S \sim \nu_S \sim \pTcut \,.
\eea
Since the only possible large logarithms will be of the form in Eq.~(\ref{logs}),
once we resum them, all of the large logarithms in Eq.~({\ref{FO}}) will be resummed to give $\sigma_{res.}$
in Eq.~({\ref{sumFO}}).

Although we have discussed the structure of the resummed cross section of the 0-jet bin in gluon-fusion, the results for the 1-jet bin and for the $VH$ process are very similar.  For the $W^{+}H$ cross section in the 0-jet bin, the singular contribution is factorized in the same way as in Eq.~(\ref{fact.}).  The only differences are the different virtual corrections to the two processes resulting in different hard functions, and the replacement of the gluon beam function needed in gluon fusion with the quark beam function.  For gluon fusion in the 1-jet bin, the factorization theorem contains an additional jet function describing collinear radiation within the final-state jet.  For more details we refer the reader to Refs.~\cite{Li:2014ria} and~\cite{Liu:2012sz} for these processes, respectively.

\subsection{Matching the resummed result with fixed-order}

The factorization theorem describes only the singular part of the cross section,
turning $\sigma_{sing.}$ into $\sigma_{res.}$ by renormalization-group evolution. To obtain the full prediction 
we must include the $\sigma_{\rm ns}$ term. By comparing Eq.~(\ref{FO}) and Eq.~(\ref{sumFO}), we 
find
\bea\label{matching}
\sigma_0^{{\rm resum}+{\rm FO}}(\pTcut) = 
\sigma_{res.} - \sigma_{sing.} + \sigma_0^{\rm FO} \,,
\eea
where $\sigma_{sing.}$ is obtained by expanding $\sigma_{res.}$ to the same order in $\alpha_s$ as $\sigma_0^{\rm FO}$.  The cross section in Eq.~(\ref{matching}) satisfies the following properties:
\bea
&&\text{for}\,  \pTcut \to 0, \quad \sigma_0^{{\rm resum}+{\rm FO}}(\pTcut)  \to \sigma_{res.} \,; \nn \\
&&\text{for}\,\pTcut \to m_H, \quad \sigma_0^{{\rm resum}+{\rm FO}}(\pTcut)  \to \sigma_0^{\rm FO} \,; \nn \\
&&\text{for}\,\pTcut \gg m_H, \quad \sigma_0^{{\rm FO}}(\pTcut) =  \sigma^{\rm FO}_{\rm inclusive} \,; \nn\\
&&\text{for}\,\pTcut \gg m_H, \quad \sigma_0^{{\rm resum}+{\rm FO}}(\pTcut) 
\neq
 \sigma^{\rm FO}_{\rm inclusive} \,. 
\eea
This last feature is problematic, since we must demand that when $\pTcut$ is large enough, the jet-vetoed cross section $\sigma_0(\pTcut)$ reproduces the inclusive fixed-order result.  To enforce this, we use the idea of profile scales~\cite{Abbate:2010xh,Ligeti:2008ac}, which smoothly merge the separate hard, soft, and beam scales introduced previously into a single scale as $\pTcut$ becomes of the same order as the hard scale:
\bea
\mu_i(\pTcut) \to \mu_{\rm FO} \,, \quad \quad \nu_i(\pTcut) \to \mu_{\rm FO} \,,
\eea
for $i = H, B, S$.  This reduces the RG-evolutions factors $U_F$ in Eq.~(\ref{Ufac}) for these functions to unity, and the resummed cross section reduces to the singular cross section, so that $\sigma_0^{{\rm resum}+{\rm FO}}(\pTcut) \to \sigma_0^{\rm FO}$ in Eq.~(\ref{matching}).  A detailed discussion of profile scales is given in Ref.~\cite{Stewart:2013faa}.  In our study we obtain the fixed-order cross sections for the Higgs+0-jet at NNLO from~\cite{Anastasiou:2004xq,Anastasiou:2005qj,Catani:2007vq,Catani:2008me} and the Higgs+1-jet process at NLO from MCFM~\cite{Campbell:2010ff}.  The fixed-order results for the $W^{+}H$ process are obtained from a modified version of FEWZ~\cite{Melnikov:2006kv,Gavin:2010az}, as described in Ref.~\cite{Li:2014ria}.

\subsection{Imaginary matching scales and $\pi^2$ resummation}
\label{sec:pisq}

The hard functions for the gluon-fusion and $VH$ processes considered here contain logarithms of the following form:
\bea
H(m_H,\mu) \supset \log^2\frac{-Q - i0}{\mu} \to \log^2\frac{Q}{\mu} - \pi^2  ,
\eea
where $Q$ is the relevant hard scale of the process.  We can extend the resummation of the logarithms $\log\frac{m_H}{\mu} $ to include the related $\pi^2$ terms by the scale choice $\mu_H = -i |\mu_H|$.  This resummation has been extensively studied in the literature~\cite{Parisi:1979xd,Sterman:1986aj,Magnea:1990zb}, and has been shown to improve the perturbative convergence of the inclusive gluon-fusion cross section~\cite{Ahrens:2008qu,Ahrens:2008nc}.  This $\pi^2$ resummation modifies the resummed cross section $\sigma_{res.}$ of Eq.~(\ref{sumFO}) in the following way:
\bea
\sigma_{res.+\pi^2} = 
 \sigma_0^{(0)}\left[1 + \alpha_s  \tilde{c}_0^{(1)} \right] 
\exp\left[\alpha_s \left( c^{(1)}_2L^2 + c^{(1)}_1L  \right) \right] 
\exp\left[\frac{\alpha_s C_A \pi }{2} + \cdots \right] \,.
\eea
The $\tilde{c}$ in the coefficient of the exponential is different from the $c$ in in Eq.~(\ref{sumFO}) by the $\pi^2$ term we have resummed:
\bea
\tilde{c}_0^{(1)} = c_0^{(1)} - \frac{\alpha_sC_A \pi}{2}  \,.
\eea
The $\pi^2$ terms being resummed must be subtracted from $c$ to avoid double counting. 

Unlike the logarithms of $\pTcut/m_H$ which should become less important 
when $\pTcut$ approaches the hard scale, the $\pi^2$ terms arise from the virtual corrections and act as a large $K$-factor to the fixed-order inclusive cross section. Therefore, we must keep the $\pi^2$ resummation turned on even for large $\pTcut$ 
to account for the large constant corrections from higher orders. To do so, we have to 
carefully subtract out the $\pi^2$ terms in the fixed-order cross section to avoid double counting.  This is done via the following matching:
\bea\label{matchpisq}
\sigma_0^{{\rm resum}+\pi^2 + {\rm FO}}(\pTcut) 
&=& \left[\sigma_{res.} - \sigma_{sing. L} + 
(\sigma_0^{\rm FO} - \sigma_{sing.\pi^2} - \sigma_{{\rm ns},\pi^2})  \right] \exp\left[\frac{\alpha_s C_A \pi }{2} + \cdots \right] \, \nn \\
&=& \left[\sigma_{res.} - \sigma_{sing.} + 
\sigma_0^{\rm FO}  - \sigma^{(1)}_{{\rm ns}}\frac{\alpha_sC_A\pi}{2}  \right] \exp\left[\frac{\alpha_s C_A \pi }{2} + \cdots \right] \,.
\eea
In the first line, $\sigma_{res.}$ means we only resum the $\pTcut$ logarithms.  $\sigma_{sing. L} $ is the singular term containing only $\pTcut$ logarithms, while $\sigma_{sing. \pi^2} $ is the singular term that contains only the $\pi^2$ terms. The full
singular term is given by the sum of these two which is why we get $\sigma_{sing.}$ in the second
line.  We also subtract the $\pi^2$ terms $\sigma_{{\rm ns},\pi^2}$ in the non-singular term $\sigma_{{\rm ns}}$.  These come from the interference between
the non-singular terms and the virtual corrections.  This contribution first appears at ${\cal O}(\alpha_s^2)$, and is given by
$\sigma^{(1)}_{{\rm ns}}\alpha_sC_A\pi/2 $, with
\bea
\sigma_{\rm ns}^{(1)} = \sigma^{\rm FO,{(1)}} - \sigma_{sing.}^{(1)} \,.
\eea
From the first line of Eq.~(\ref{matchpisq}), we see that when $\pTcut$ is large,
\bea
\sigma_0^{{\rm resum}+\pi^2 +{\rm FO}} \to \tilde{\sigma_0}^{\rm FO}
\exp\left[\frac{\alpha_s C_A \pi }{2} + \cdots \right] \,,
\eea
where $\tilde{\sigma_0}^{\rm FO}$ is the FO cross section with $\pi^2$ terms suitably subtracted.  This is the desired expression, as the large constant $\pi^2$ corrections to the fixed-order cross section are still resummed when the $L$-resummation is turned off.

\section{Numerical results}
\label{sec:numerics}

We now present numerical results for Higgs production in 33 and 100 TeV $pp$ collisions.  We follow the theoretical formalism presented in the previous section.  We show results for gluon-fusion production in association with 0, 1, and 2 or more jets.  This division into bins of jet multiplicity is used in the current LHC analyses in the $WW$ channel.  We also present results for $W^+ H$ production in the 0-jet bin, as suggested to measure the Higgs coupling to bottom quarks~\cite{Butterworth:2008iy}.  The results for $W^- H$ production are very similar. Although we explained the theoretical framework using Higgs production in the 0-jet bin of gluon fusion, the structure of the results is similar for all channels.  We refer the reader to Refs.~\cite{Boughezal:2013oha,Li:2014ria} for more details.  For the fixed-order cross sections, we use the MSTW parton distributon functions at the same order in perturbation theory as the partonic cross section~\cite{Martin:2009iq}.  For the resummation-improved gluon-fusion and  VH cross sections, we use the NNLO MSTW distribution functions.  

\begin{figure}[htbp]
\begin{center}
\includegraphics[width=0.75\textwidth,angle=0]{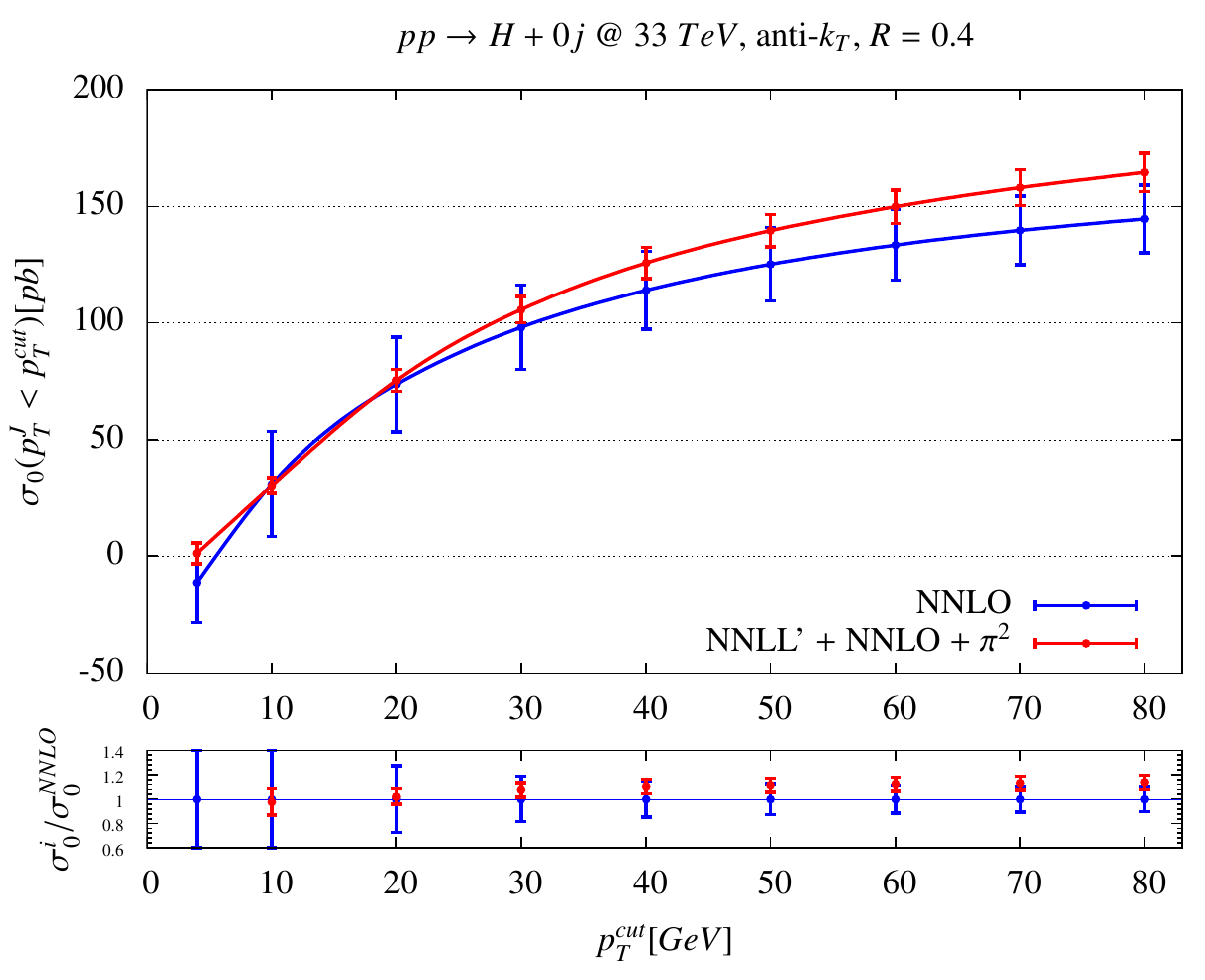}
\end{center}
\caption{Shown is the fixed-order cross section for the H$+0$-jet cross section in blue and the resummation-improved result in red as a function of the $p_T^{cut}$ for a future $33$ TeV pp machine.  The relative changes induced by the resummation with respect to the fixed-order results are shown in the lower panels.}
\label{fig:H0j33}
\end{figure}

\begin{figure}[htbp]
\begin{center}
\includegraphics[width=0.75\textwidth,angle=0]{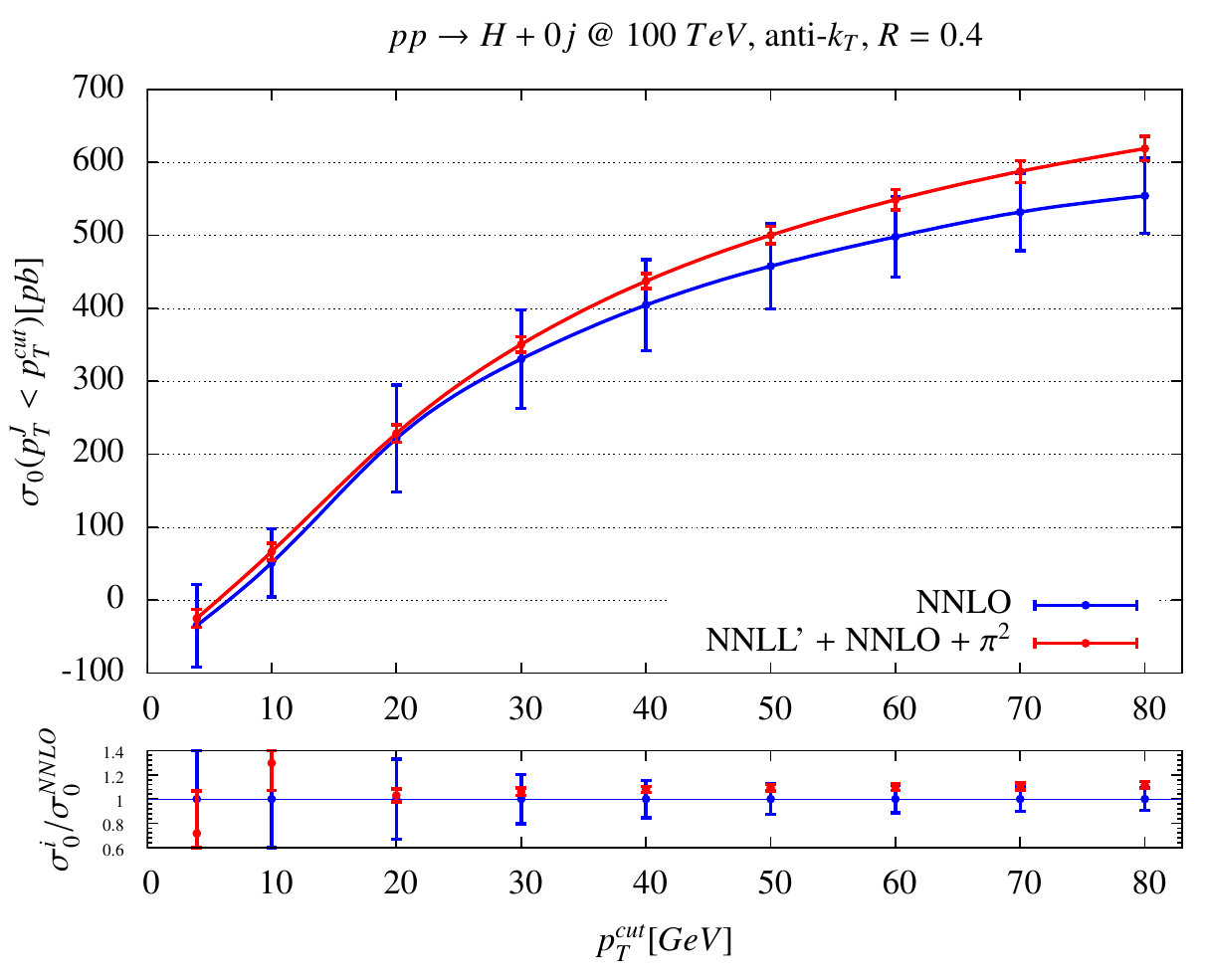}
\end{center}
\caption{Shown is the fixed-order cross section for the H$+0$-jet cross section in blue and the resummation-improved result in red as a function of the $p_T^{cut}$ for a future $100$ TeV pp machine.  The relative changes induced by the resummation with respect to the fixed-order results are shown in the lower panels.}
\label{fig:H0j100}
\end{figure}

\subsection{Results for gluon-fusion in 33 and 100 TeV proton-proton energies   }

We begin by showing predictions for Higgs production in the 0-jet, 1-jet, and inclusive 2-jet bins.  We compare the highest-order fixed-order predictions with those improved by resummation of jet-veto logarithms.  For Higgs production in the 0-jet bin, we can resum the jet-veto logarithms through the $\text{NNLL}^{\prime}+\text{NNLO}$ order, using the logarithm-counting scheme presented in Ref.~\cite{Berger:2010xi}.  We further choose an imaginary matching scale in order to resum enhanced $\pi^2$ terms, as explained in Section~\ref{sec:pisq}.  The fixed-order cross section is known through NNLO~\cite{Anastasiou:2004xq,Anastasiou:2005qj,Catani:2007vq,Catani:2008me}.  For Higgs production in the exclusive 1-jet bin, we resum the global jet-veto logarithms through the $\text{NLL}^{\prime}+\text{NLO}$ order.  In order to combine with the 0-jet bin, a similar imaginary matching scale is chosen, and partial NNLO results for the 1-jet cross section~\cite{Boughezal:2013uia} are implemented.  This procedure is described in detail in Ref.~\cite{Boughezal:2013oha}.  The effects from non-global logarithms were discussed in Ref.~\cite{Liu:2013hba}, and were found to be small.  The fixed-order cross section is known through NLO, and is implemented in MCFM~\cite{Campbell:2010ff}.  The inclusive 2-jet cross section is also known through NLO, and is implemented in MCFM~\cite{Campbell:2010cz}.  It is currently not possible to directly resum the large logarithms in the inclusive 2-jet bin due to the presence of numerous scales.  However, unitarity allows us to write this cross section in terms of the total cross section and the exclusive 0-jet and 1-jet cross sections that we can renormalization-group (RG)-improve:
\begin{equation}
\sigma_{\geq 2} = \sigma_{tot}-\sigma_0-\sigma_1.
\end{equation}  
We compare the result obtained in this way against that obtained at fixed-order.  For the central values, we choose the scale to be $\mu = m_H$. 
This is consistently chosen to be the same for the total cross section, the exclusive  0-jet and 1-jet cross sections as well as the $ \geq 2j$ result, unless otherwise stated.

The 0-jet cross sections as a function of $p_T^{cut}$ are shown in Figs.~\ref{fig:H0j33}~and~\ref{fig:H0j100} for 33 and 100 TeV $pp$ collisions.  Since the jet thresholds at future high-energy hadron colliders are unknown, we have allowed $p_T^{cut}$ to vary from below its current LHC value of approximately $25-30$ GeV up to 80 GeV.  As indicated in the plots, anti-$k_T$ jets with $R=0.4$ are used.  The resummation-improved predictions are higher than the fixed-order ones by approximately $5-10\%$ for values of $p_T^{cut}$ above 30 GeV at both collider energies.  This is driven by the $\pi^2$ resummation accomplished by the imaginary matching scale chosen.  The breakdown of fixed-order perturbation theory is apparent for lower values of $p_T^{cut}$. The uncertainties after RG-improvement are decreased by more than a factor of two over the entire kinematic range.  However, the general impact of the jet-veto logarithms in the 0-jet bin are similar at 33 and 100 TeV to what was found at lower energies~\cite{Becher:2012qa,Tackmann:2012bt,Becher:2013xia,Stewart:2013faa,Banfi:2012yh,Banfi:2012jm}.  Gluon-fusion Higgs production is dominated by values of partonic scattering energies $\sqrt{\hat{s}} \sim m_H$ because of the rapid fall-off of the gluon luminosity as Bjorken-$x$ is increased.  The relevant hard scale in the jet-veto logarithms is therefore set by $m_H$ at these collider energies.  Also, the $\pi^2$-resummation accounts for a large amount of both the shift in central value and the decreased uncertainty, and this has no dependence on the collider energy. The change in going from fixed-order to resummation therefore differs little in these higher energy collisions.

\begin{figure}[htbp]
\begin{center}
\includegraphics[width=0.75\textwidth,angle=0]{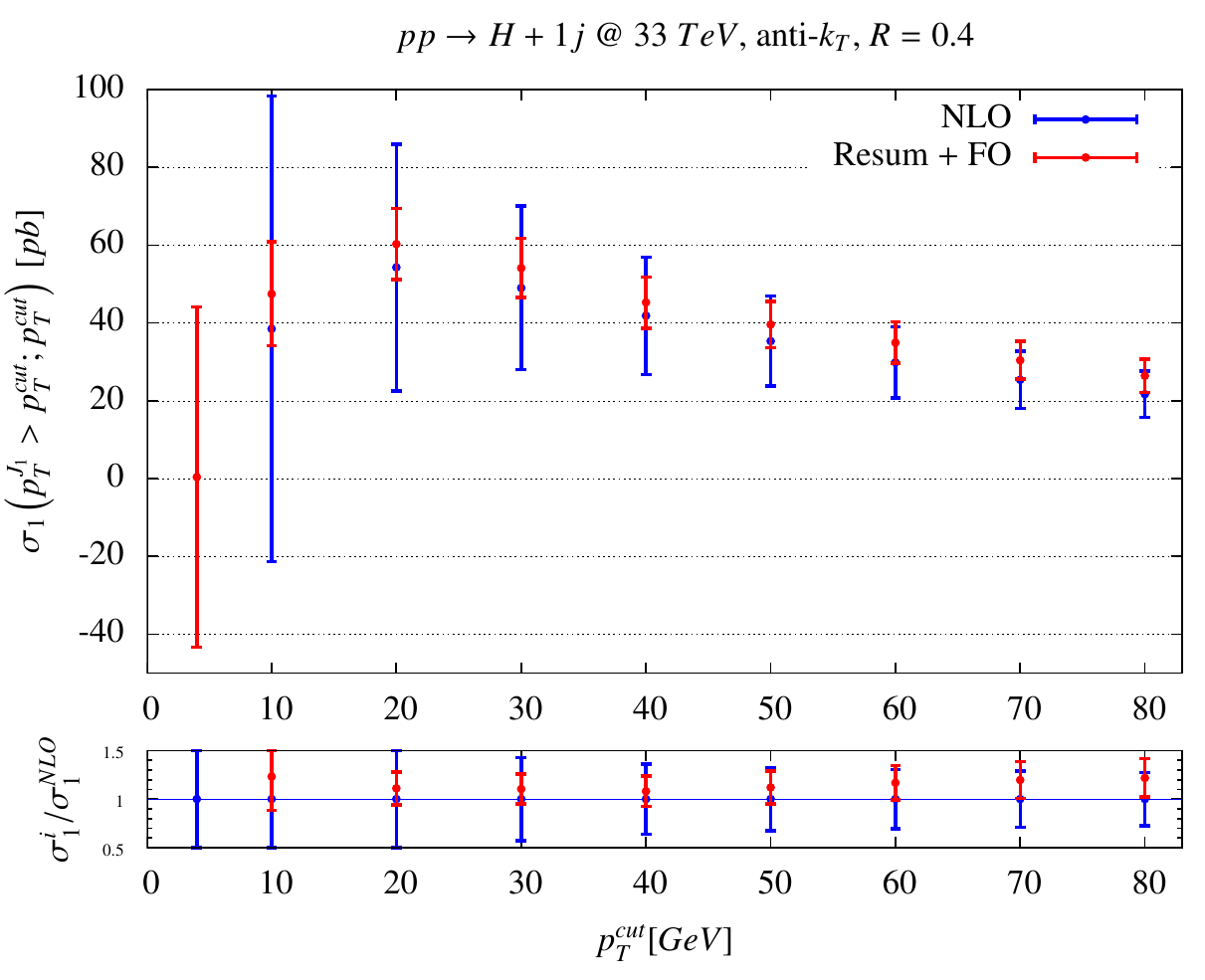}
\end{center}
\caption{Shown is the fixed-order cross section for the H$+1$-jet cross section in blue and the resummation-improved result in red as a function of the $p_T^{cut}$ for a future $33$ TeV pp machine.  The relative changes induced by the resummation are shown in the lower panels.}
\label{fig:H1j33}
\end{figure}

\begin{figure}[htbp]
\begin{center}
\includegraphics[width=0.75\textwidth,angle=0]{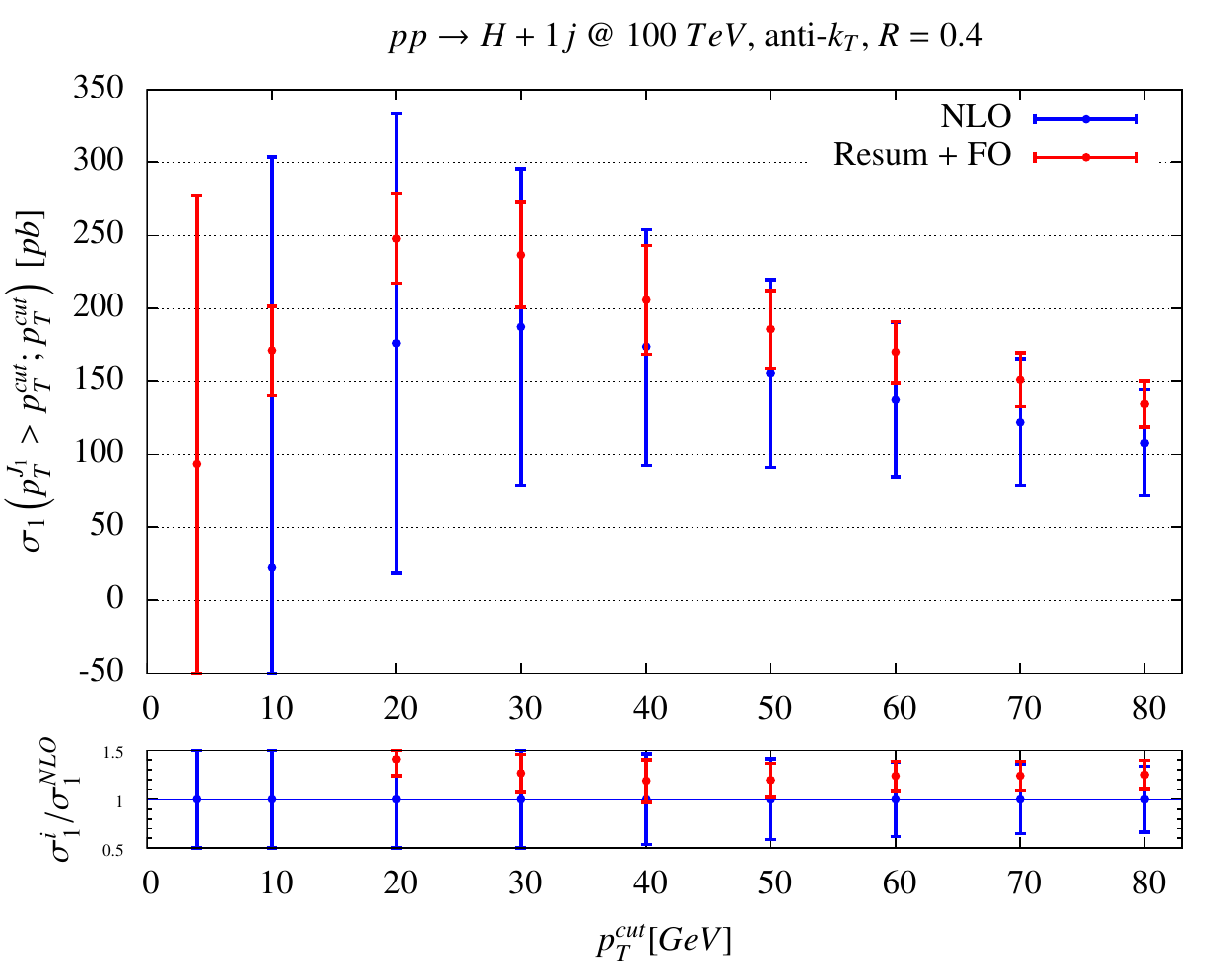}
\end{center}
\caption{Shown is the fixed-order cross section for the H$+1$-jet cross section in blue and the resummation-improved result in red as a function of the $p_T^{cut}$ for a future $100$ TeV pp machine.  The relative changes induced by the resummation are shown in the lower panels.}
\label{fig:H1j100}
\end{figure}

The results for the 1-jet bin are shown in Figs.~\ref{fig:H1j33}~and~\ref{fig:H1j100} for both collider energies.  The shifts in the central value when incorporating the resummation improvement are slightly larger than the 0-jet results.  At 33 TeV they lead to a 10\% increase in the predicted cross section over a wide range of $p_T^{cut}$, while at 100 TeV the predicted cross section is increased by up to 30\%.  The theoretical uncertainties are more significantly reduced as the collider energy is increased.  The fixed-order estimated errors grow to over $\pm 50\%$ for $p_T^{cut} \leq 40$ GeV in 100 TeV collisions.  The resummation uncertainties remain at or below $\pm 20\%$ for all collider energies and for all relevant $p_T^{cut}$ values.  The relevant logarithms for the 1-jet cross section are $\text{ln}(p_{TJ}/p_T^{cut})$.  Although the 1-jet bin receives most of its contribution from the low-$p_{TJ}$ region, there is still a significant tail at high-$p_{TJ}$.  This tail contributes a large amount of the scale variation at fixed-order even though it is sub-dominant in the rate~\cite{Liu:2013hba}.  At higher collider energies there is more contribution from this high-$p_T$ region due to the increased phase space available.  Since the logarithms become very large at high jet $p_T$, there is a relatively larger reduction in the theoretical error obtained by resuming these terms as the collider energy is increased.

\begin{figure}[htbp]
\begin{center}
\includegraphics[width=0.75\textwidth,angle=0]{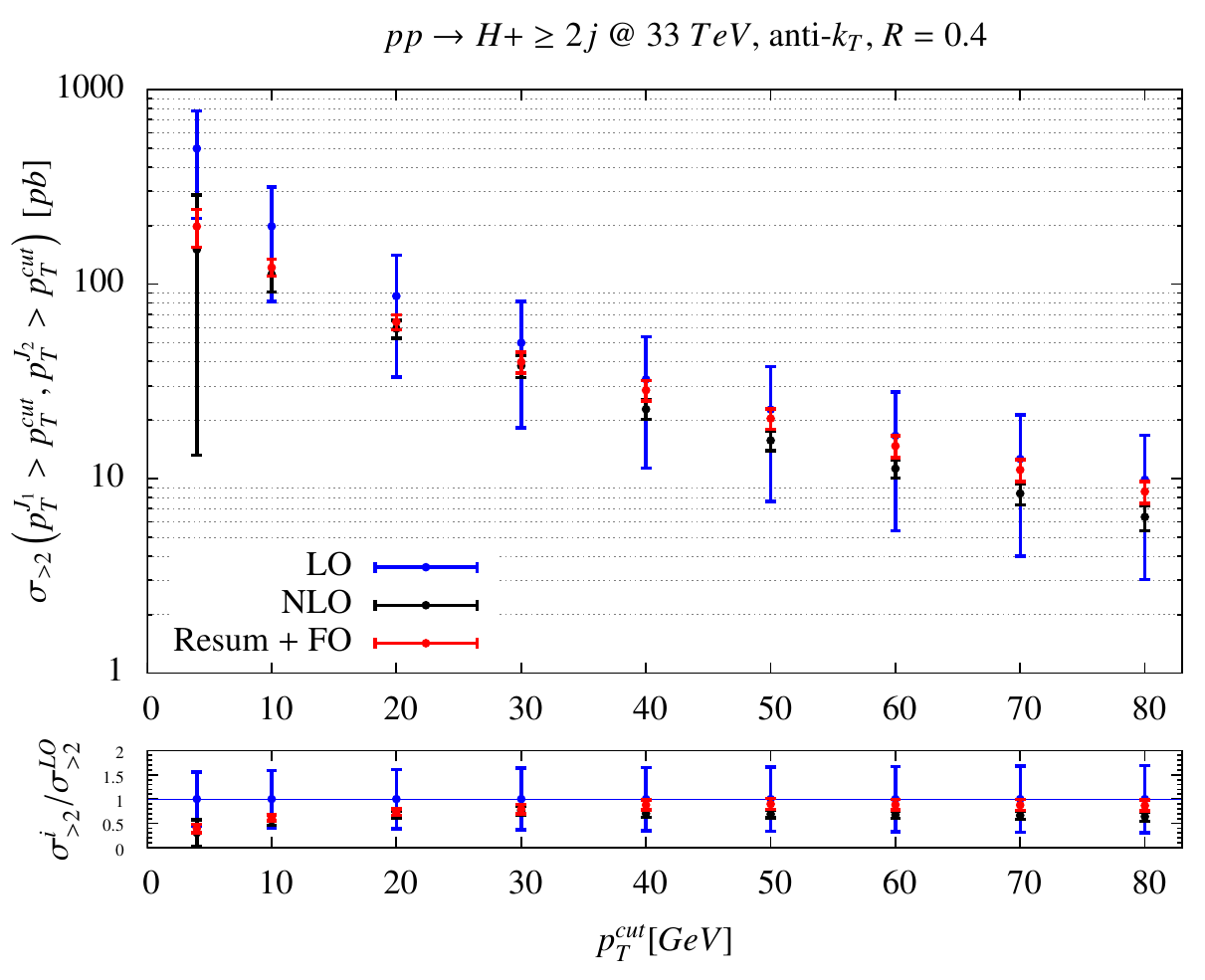}
\end{center}
\caption{Shown is the fixed-order cross section for the H$+2$-jet cross section in blue and the resummation-improved result in red as a function of the $p_T^{cut}$ for a future $33$ TeV pp machine.  The relative changes induced by the resummation are shown in the lower panels.  Both the LO and NLO fixed-order results have been included.}
\label{fig:H2j33}
\end{figure}

\begin{figure}[htbp]
\begin{center}
\includegraphics[width=0.75\textwidth,angle=0]{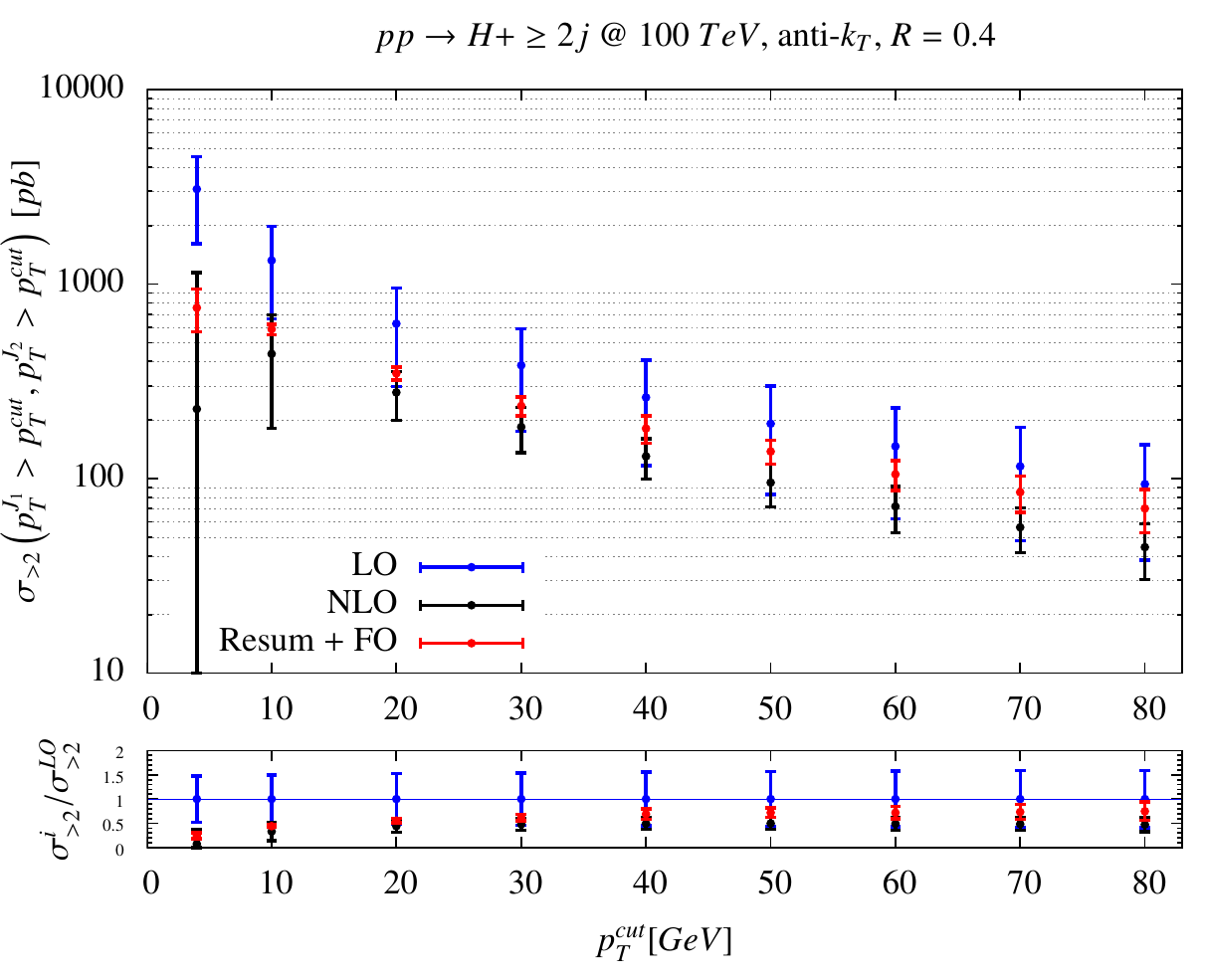}
\end{center}
\caption{Shown is the fixed-order cross section for the H$+2$-jet cross section in blue and the resummation-improved result in red as a function of the $p_T^{cut}$ for a future $100$ TeV pp machine.  The relative changes induced by the resummation are shown in the lower panels.  Both the LO and NLO fixed-order results have been included.}
\label{fig:H2j100}
\end{figure}

Finally, we compare the resummation-improved predictions for the inclusive 2-jet bin against the fixed-order result.  Although the jet-veto logarithms in the inclusive 2-jet cross section cannot be directly resummed, a resummation-improved prediction can be obtained by using the total cross-section constraint: $\sigma_{\geq 2} = \sigma_{tot}-\sigma_0-\sigma_1$.  There are several choices for which fixed-order result to use.  The NLO calculation is known and incorporated into MCFM~\cite{Campbell:2010cz}.  However, combining this calculation with the NNLO total cross section and the NLO inclusive 1-jet would lead to a mismatch in the order of $\alpha_s$ used for each cross section; the total and inclusive 1-jet results are through ${\cal O}(\alpha_s^2)$, while the NLO 2-jet cross section is through ${\cal O}(\alpha_s^3)$.  Since the exclusive 1-jet cross section is obtained from the difference $\sigma_1 = \sigma_{\geq 1} - \sigma_{\geq 2}$ in the fixed-order ST method, there may potentially be an incorrect estimate of the large logarithms appearing at ${\cal O}(\alpha_s^3)$ if the NLO inclusive 2-jet result is used.  Currently, the ATLAS collaboration uses the LO cross section in their $WW$ analysis, while CMS uses the NLO result.  We compare against both the LO and NLO fixed-order results.

We show results for 33 and 100 TeV energies in Figs.~\ref{fig:H2j33} and~\ref{fig:H2j100}.  There are several issues to note from these plots.  The LO fixed-order uncertainties are over $\pm 50\%$ at all energies.  This is drastically reduced in both the NLO and resummation-improved results, where it is instead at the level of $\pm 10-15\%$.  The mismatch between the central values predicted by LO and those obtained at NLO and with resummation grows with collider energy, reaching 50\% at a 100 TeV machine.  The central values at NLO and those obtained by RG-improvement exhibit a better agreement, with discrepancies reaching $\pm 20\%$ at 100 TeV.  Fixed-order perturbation theory does a rather good job of predicting the cross section central values over a very wide range of collider energy and even down to low values of 
$p_T^{cut} \sim 20-30$ GeV.

\begin{figure}[htbp]
\begin{center}
\includegraphics[width=0.75\textwidth,angle=0]{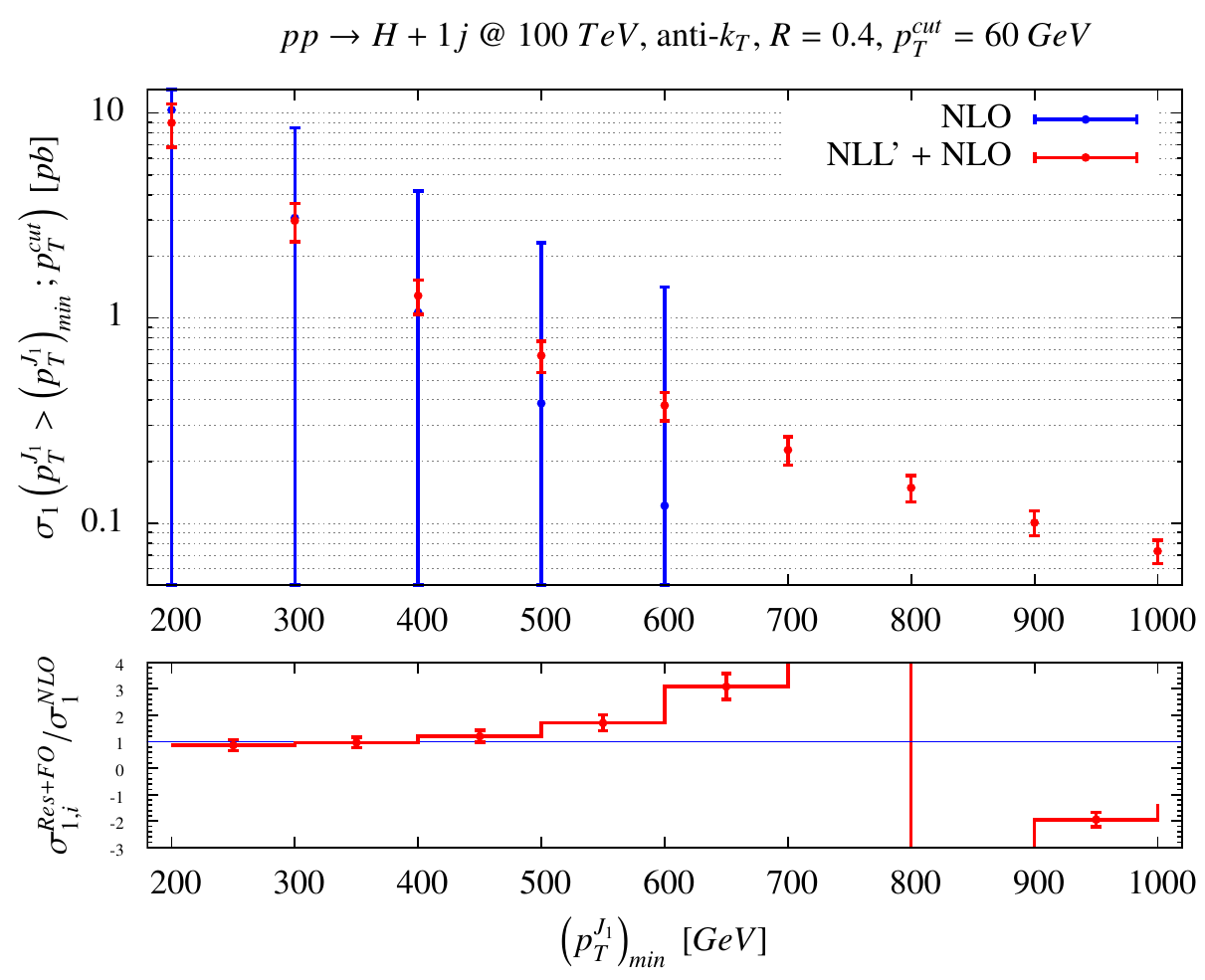}
\end{center}
\caption{Shown is the 1-jet cross section in both fixed-order and RG-improved perturbation theory as a function of a lower cut on the $p_T$ of the jet.  The high-$p_T$ region with $(p_{T}^{J_1})_{min} > 200$ GeV is shown. The relative deviation between the two is shown in the lower inset.  }
\label{fig:H1j100boost}
\end{figure}

We finally show one more result which clearly demonstrates the importance of comparing other predictions against RG-improved perturbation theory at a future hadron collider.  These machines will allow new kinematic regions to be explored, permitting studies of the Higgs at energies and momenta far beyond what the LHC can produce.  One interesting observable is the study of the Higgs recoiling against a very high-$p_T$ jet.  This has been suggested as an interesting probe of possible beyond-the-SM effects~\cite{Azatov:2013xha}.  One could imagine the need to consider exclusive jet bins as a way to separate the gluon-fusion and vector-boson fusion components of the cross section.  We show in Fig.~\ref{fig:H1j100boost}
the 1-jet cross section as a function of a lower cut on the transverse momentum of the jet.  We consider cuts ranging from 200 to 1000 GeV in accordance with the large kinematic range available at a 100 TeV machine.  We have chosen $\pTcut = 60$ GeV.  For the fixed-order result we have used the central scale choice $\mu = 2 p_T^{J_1,min}$; the NLO result becomes negative for the fixed scale $\mu=m_H$.  We note that the uncertainties on the fixed-order result are enormous.  They make the NLO results consistent with zero within the estimated uncertaintiesover the entire kinematic range studied.  The fixed-order central value becomes negative when $p_T^{J1,min} \approx 600$ GeV, rendering fixed-order perturbation theory unusable.  In contrast, the RG-improved result exhibits small scale dependence and sensible central values over the entire range studied.  Resummation is mandatory when extending exclusive jet binning into new kinematics regimes at future facilities.

\subsection{Results for $W^{+}H$ production in 33 and 100 TeV proton-proton collisions}

We now consider the $W^{+} H$ process in the 0-jet bin. We focus on the boosted regime, in which the Higgs is produced at high transverse momentum.  In that region the two bottom quarks coming from the Higgs decay are collimated, creating a ``fat jet" which can be searched for experimentally.  This is a promising channel in which to measure the bottom-quark coupling to the Higgs.  A jet veto is imposed in the suggested analysis to remove backgrounds from $t\bar{t}$ production~\cite{Butterworth:2008iy}.  By going to the boosted region while introducing a jet veto, large logarithms of the approximate form 
$\text{ln}(p_T^W/p_T^{cut})$ are obtained; we study their effect on the $W^{+}H$ cross section here.  In this section we set the anti-$k_T$ jet radius to $R=1.2$ to mimic the fat jet suggested in the analysis.  We compare the RG-improved cross section through the $\text{NNLL}^{\prime}+\text{NNLO}$ order~\cite{Li:2014ria} with the NNLO fixed-order result~\cite{Ferrera:2011bk}.

\begin{figure}[htbp]
\begin{center}
\includegraphics[width=0.75\textwidth,angle=0]{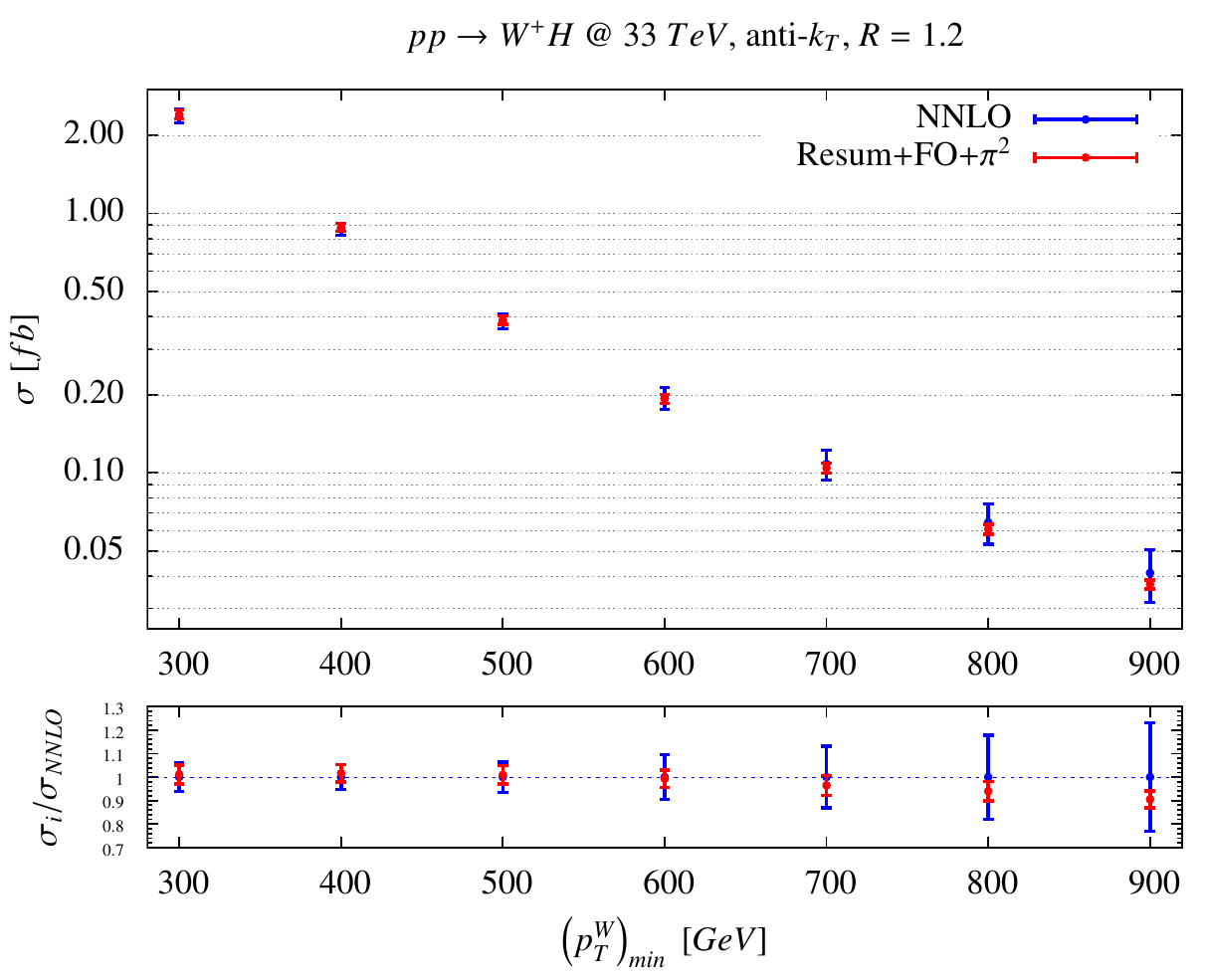}
\end{center}
\caption{Shown is the NNLO fixed-order cross section for the $W^{+}H$ process and the resummation improved result in red as a function of $(p_{T}^{W})_{min}$ for a future $33$TeV pp machine.  The relative changes induced by the resummation are shown in the lower panels.}
\label{fig:WH33}
\end{figure}

We begin by showing the result for $W^{+} H$ production in 33 TeV $pp$ collisions in Fig.~\ref{fig:WH33}.  These results are the cumulant distributions, with a cut on the transverse momentum of the $W$-boson of $p_T^W > p_{T,min}^W$.  We show predictions up to $p_{T,min}^W = 900$ GeV, for bins of 100 GeV.  The highest cumulant bin will contain roughly 50 events assuming 100 fb$^{-1}$ of integrated luminosity, suggesting that this region will be statistically observable at future machines.  The gradual increase in the size of the jet-veto logarithms is apparent at high-$p_T^W$.  Deviations of 10\% from the NNLO prediction are seen at high transverse momenta, although there is a good agreement between the fixed-order and resummation-improved predictions over the lower $p_{T,min}^W$ range.  The theoretical uncertainties are greatly improved by including the resummation.  For $p_{T,min}^W = 600$ GeV, they are reduced by a factor of two, while for the upper edge $p_{T,min}^W = 900$ GeV they are reduced by a factor of four.

\begin{figure}[htbp]
\begin{center}
\includegraphics[width=0.75\textwidth,angle=0]{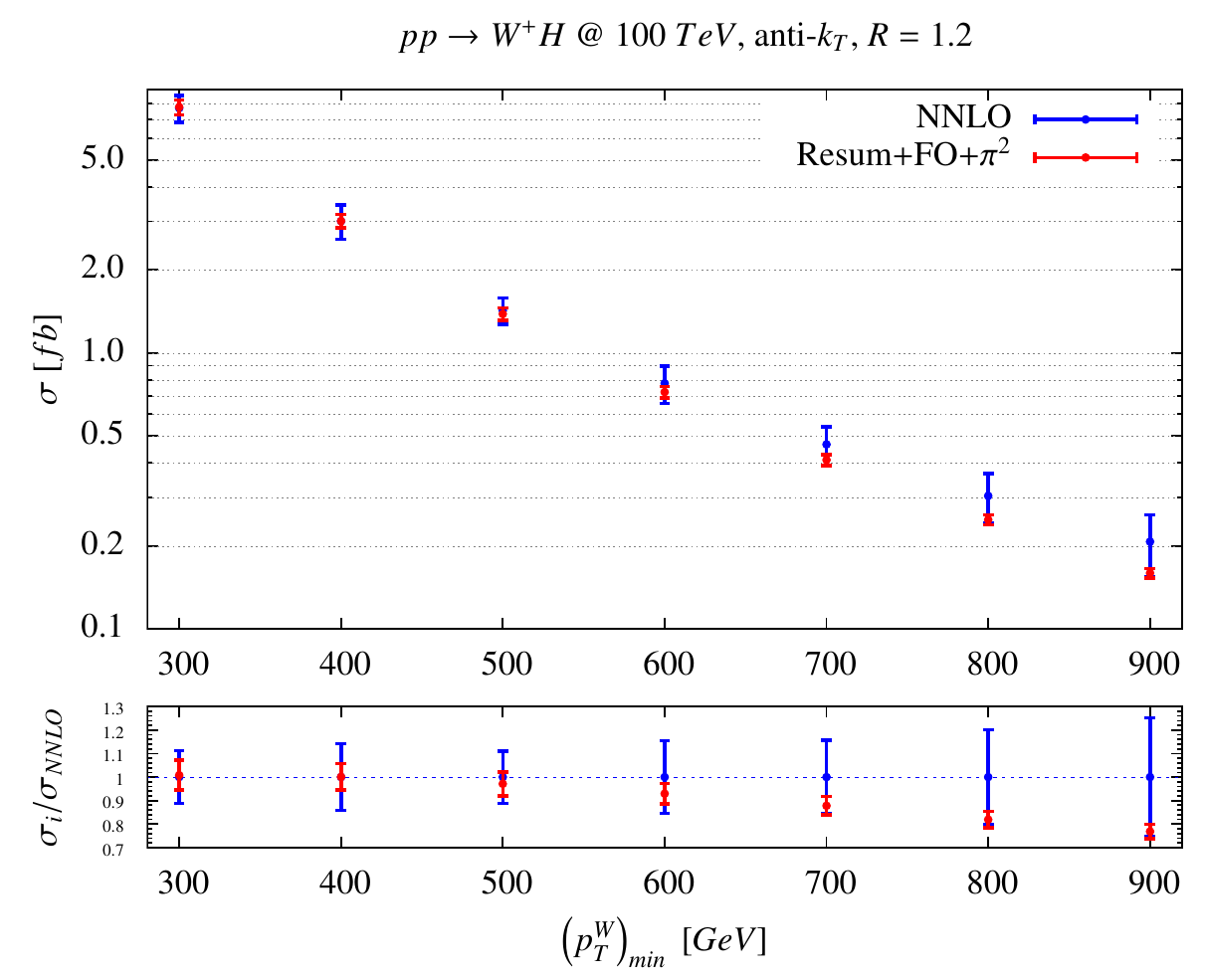}
\end{center}
\caption{Shown is the NNLO fixed-order cross section for the $W^{+}H$ process and the resummation improved result in red as a function of $p_{T,min}^{W}$ for a future $100$TeV pp machine.  The relative changes induced by the resummation are shown in the lower panels.}
\label{fig:WH100}
\end{figure}

The result for 100 TeV $pp$ collisions is shown in  Fig.~\ref{fig:WH100}.  Although we keep the $p_{T,min}^W$ range the same as before, there is increased phase-space available for high-energy partonic collisions at 100 TeV, leading to a more pronounced difference between the fixed-order result and the resummation-improved one.  The deviations reach 20\% at high $p_{T,min}^W$, and are 10\% already at $p_{T,min}^W = 600$ GeV.  The reduction in the theoretical uncertainty after incorporating the resummation is dramatic.  In the highest region of $p_{T,min}^W$, the uncertainty decreases from more than $\pm 20$\% at NNLO to under $\pm 5$\%, indicating the utility of resummation in taming the large logarithms appearing in the high-energy scatterings.

\begin{figure}[htbp]
\begin{center}
\includegraphics[width=0.75\textwidth,angle=0]{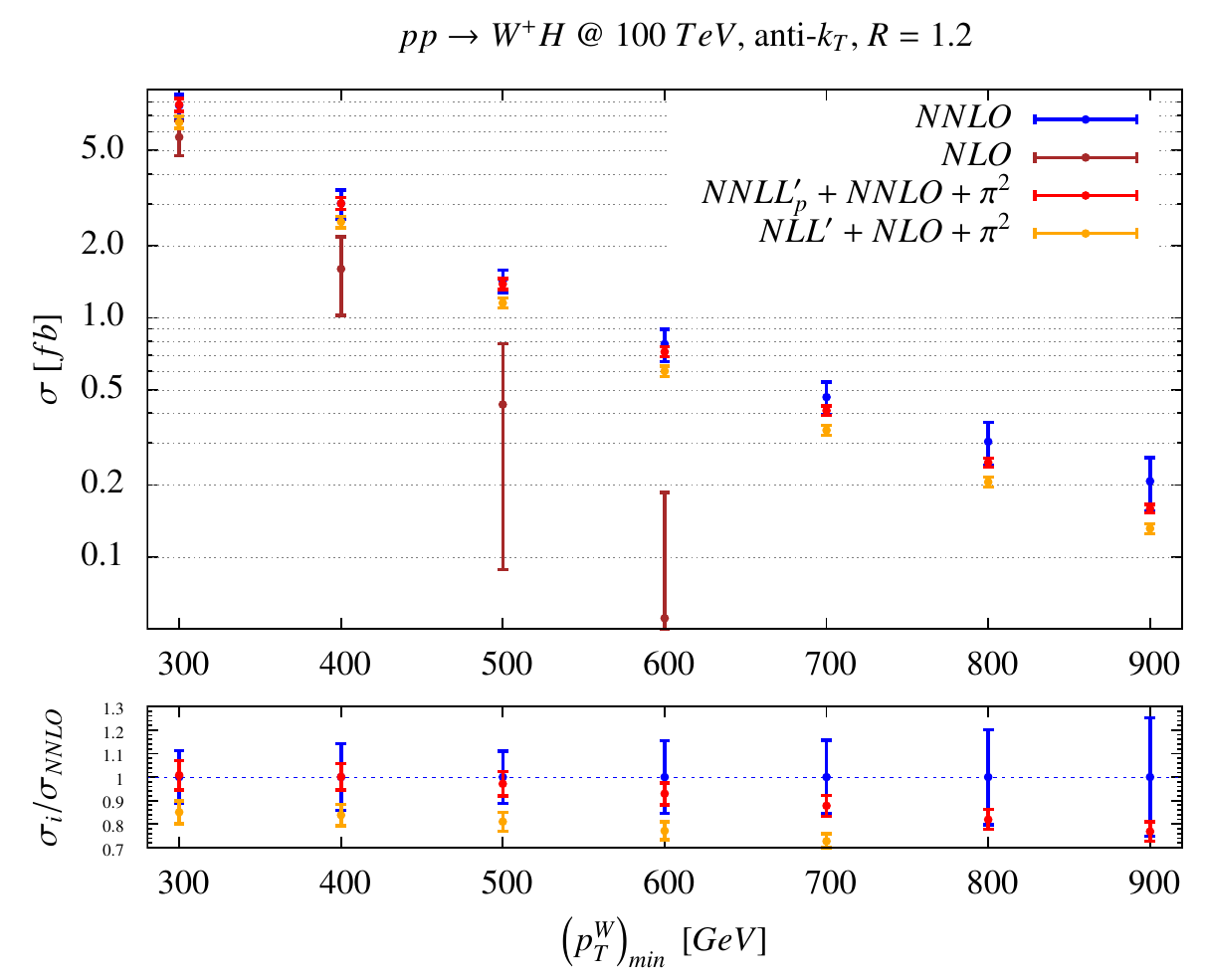}
\end{center}
\caption{Shown are the NLO, NNLO, $\text{NLL}^{\prime}+\text{NLO}$ and $\text{NNLL}^{\prime}+\text{NNLO}$ cross sections for the $W^{+}H$ process at a future $100$TeV pp machine.  The relative deviations from NNLO are shown in the lower panels.}
\label{fig:$W^{+}H$100convg}
\end{figure}

One final issue we wish to address is the convergence of both the fixed-order perturbative expansion and the RG-improved one.  Although the relative deviations between the two reach only 20\% in the kinematic region studied, this does not tell the entire story.  The RG-improved framework is dramatically more stable than fixed-order at these energies.  To demonstrate this, we consider the vetoed $W^{+}H$ cross section also at one perturbative order below the highest we can obtain.  This means that we study the fixed-order expansion at both NLO and NNLO, and the RG-improved result at both $\text{NLL}^{\prime}+\text{NLO}$ and $\text{NNLL}^{\prime}+\text{NNLO}$.  The results are shown in Fig.~\ref{fig:$W^{+}H$100convg}.  The fixed-order perturbative expansion shows no sign of convergence as the order is increased from NLO to NNLO.  For $p_{T,min}^W \approx 500$ GeV the NLO result is a factor of three less than the NNLO one.  It becomes negative at $p_{T,min}^W \approx 600$ GeV.  In contrast, the corrections $W^{+}H$ when increasing the order of resummation-improved perturbation theory are far more modest, on the order of $20-30\%$ even at high $p_{T,min}^W$.  Without the resummation of the jet-veto logarithms, no reliable estimate of this cross section can be obtained in high-energy collisions. 

\section{Conclusions}
\label{sec:conclusion}

In this manuscript we have studied properties of Higgs boson production in bins of exclusive jet multiplicity at future 33 and 100 TeV proton-proton collisions.  The increase in collider energy permits partonic scattering at very high $\sqrt{\hat{s}}$, potentially introducing very large ratios of scales and the corresponding large logarithms into theoretical predictions.  Since obtaining a detailed understanding of the Higgs will undoubtedly be a major component of the physics program at these future machines, and because jet binning has been an important part of the Higgs program at the LHC, we study the large logarithms at these higher collider energies.  We compare the best-available predictions using both fixed-order perturbation theory and resummation-improved perturbation theory, focusing on gluon-fusion Higgs production and associated $W^{+}H$ production as example processes in which the use of jet binning has previously been necessary.  We study the dependence of the predictions on the jet-veto scale, and look at a range of kinematic regions.

There are several interesting conclusions of our study, depending on the observable considered and the question asked. Fixed-order predictions taken ``out-of-the-box" agree with those obtained with RG-improvement as long as the bulk of the available phase space is considered (i.e., kinematic corners such as high-$p_T$ regions are not specifically selected).  The differences between the fixed-order cross sections and the RG-improved results are typically 20\% or less.  There are a few reasons for this result.  For Higgs production in gluon fusion, the steeply-falling gluon luminosity restricts the number of high-energy partonic scattering events in which large scale hierarchies are produced, reducing the changes in central values.  For associated WH production, the Casimir multiplying the logarithms is $C_F$ rather than the $C_A$ which occurs for gluon fusion, again reducing the changes caused by resummation. 

However, one of the major purposes of future hadron facilities is to explore new kinematic regimes in which energy scales beyond the Standard Model may manifest themselves.  These typically involve high transverse momenta, and we have studied here two motivated examples which focus on such regions: the production of a Higgs in association with a jet in the high-$p_T$ region of the jet, and the production of a boosted Higgs at high-$p_T$ in the $W^{+}H$ process.  In the first case the use of resummation is mandatory.   The fixed-order expansion breaks down at transverse momenta of a few hundred GeV, which is easily accessible at such future facilities.  The validity of the $W^{+}H$ production extends to transverse momenta of 1 TeV and beyond, due to the smaller Casimir multiplying the jet-veto logarithms for this process.

Another issue is the reliability of the perturbative expansion, usually quantified by the scale variation, or the convergence when going from one order of perturbation to a higher one.  Both measures are dramatically improved by including resummation for all observables studied.  For the gluon-fusion processes studied, the scale variation error in the 1-jet bin is reduced by more than a factor of two when the resummation is incorporated.  Even though the contribution to the rate of very high-$p_T$ jet production is suppressed by the gluon luminosity, these events contribute a large uncertainty, as previously pointed out~\cite{Liu:2013hba}. For the $W^{+}H$ associated production process, the fixed-order perturbative expansion shows no sign of convergence, with the result changing by a factor of three when going from NLO to NNLO.  The behavior is drastically improved when resummation is added.  Going from $\text{NLL}^{\prime}+\text{NLO}$ to $\text{NNLL}^{\prime}+\text{NNLO}$ leads to a correction of $20-30\%$, as expected for a well-behaved perturbative expansion.  Only within the framework of resummation can a trustworthy perurbative expansion be obtained. 

In summary, high-energy hadron colliders are an exciting possibility for the future of the high energy program.  The study of the Higgs boson will be a central aspect of study at such machines.  We have studied several aspects of the effects of QCD on the Higgs signal when bins of exclusive jet multiplicity are considered.  Our results help inform the physics studies at such future facilities, and we look forward to future extensions of this work.

\section{Acknowledgments}
\label{sec:acks}

The work of R.B. was supported by the U.S. Department of Energy, Division of High Energy Physics, under contract DE-AC02-06CH11357. The work of C.F. and X.L.  was supported by the U.S. Department of Energy, Division of High Energy Physics, under contract DE-AC02- 06CH11357 and the grants DE-FG02-95ER40896 and DE-FG02-08ER4153. The work of Y.L. was supported by the US Department of Energy under contract DE-AC02-76SF00515.
This research used resources of the National Energy Research Scientific Computing Center, which is supported by the Office of Science of the U.S. Department of Energy under Contract No. DE-AC02-05CH11231.


\end{document}